\documentstyle[twocolumn,aps,prb,epsf]{revtex}

\begin{document}
\draft
\title{Generalized Valence Bond State and Solvable Models for Spin-1/2 Systems with
Orbital degeneracy}
\author{Shun-Qing Shen}
\address{Department of Physics, the University of Hong Kong, Pokfulam, Hong Kong,
China}
\date{May 6, 2001}

\twocolumn[\hsize\textwidth\columnwidth\hsize\csname@twocolumnfalse\endcsname
\maketitle

\begin{abstract}
A spin-1/2 system with double orbital degeneracy may possess SU(4) symmetry. 
According to the group theory a global SU(4) singelt state can be expressed 
as a linear combination of all possible configurations consisting of four-site 
SU(4) singlets. Following P. W. Andersion's idea for spin 1/2 system, we propose that
the ground state for the antiferromagnetic SU(4) model is 
SU(4) resonating valence bond (RVB) state.
A short-range SU(4) RVB state is a spin and orbital liquid, and its elementary excitations 
has an energy gap. We construct a series of solvale models which ground states are short-range 
SU(4) RVB states. The results can be generalized to the antiferromagnetic SU(N) 
models.
\end{abstract}

\pacs{PACS numbers: 75.10-b, 71.70.Ej,71.27.+a}

]

Electron configurations in transition metal oxides usually have an orbital
degeneracy in additional to spin degeneracy. Strong Coulomb interaction in
these systems may produce spin systems with orbital degeneracy (for an
overview see Refs.\cite{Kugel82,Imada99}). Several coupled spin-orbital
models arise for many kinds of relevant materials. At a symmetric point the
models may possess higher symmetry, such as SU(4).\cite
{Castellani78,Rice95,Li98} Systematic study of the symmetric models may help
us to understand physical properties for realistic systems. SU(4)
spin-orbital model is a good candidate to investigate coupled spin-orbital
system. It can be solved exactly in one-dimensional case by means of Bethe
ansatz.\cite{Sutherland75} There are a lot of numerical and analytical
calculations, and most are limited in one-dimension or small clusters. Very
few rigorous results for this system are known. Oppositely we have deep
understanding on spin SU(2) systems. Some rigorous results and solvable
models are established. For instance, it was proved that the spin 1/2
antiferromagnetic periodic chain of length L has a low-energy excitation of
order 1/L.\cite{Lieb61} In the case of SU(2) system, Anderson proposed a
resonating valence bond (RVB) state as the ground state for a spin-1/2
antiferromagnet. \cite{Anderson73} \ In each configuration all spins form
spin singlet pairs, and the RVB state is composed of all possible
configurations. In fact the state is a completely general description for a
global singlet state.\cite{Ma88} His idea was applied to explain
unconventional properties of spin liquids. Some solvable models were
constructed based on the idea.\cite{Affleck87} In this paper we generalize
Anderson's RVB\ idea to a coupled spin-orbital system. We first derive
several identities for SU(4) symmetric spin-orbital system, and then prove a
rigorous statement on the SU(4) isotropic state. The state consists of SU(4)
singlets, which can be regarded as a generalized SU(4)\ RVB states. To
illustrate the idea, we construct two types of solvable models and evaluate
the ground state energies. One ground state is a short-range SU(4) RVB\
solid, and another one is a spin-orbital liquid.

We start with a Hamiltonian for a spin 1/2 system with two-fold orbital
degeneracy, which was derived by Castellani et al.\cite{Castellani78} By
neglecting the Hund's rule coupling between different on-site orbitals, the
system possesses SU(4) symmetry. The symmetric spin-orbital Hamiltonian is
expressed in terms of two sets of independent spin 1/2 operators,\cite
{Rice95} 
\begin{eqnarray}
H &=&\sum_{ij}J_{ij}\left( 2{\bf S}_{i}\cdot {\bf S}_{j}+\frac{1}{2}\right)
\left( 2\tau _{i}\cdot \tau _{j}+\frac{1}{2}\right)  \nonumber \\
&=&\sum_{ij}J_{ij}P_{ij}.  \label{model}
\end{eqnarray}
The three operators for spin, ${\bf S}_{i}^{\alpha },$ three operators for
orbital $\tau _{i}^{\beta }$, and nine operators for their direct
multiplications $2{\bf S}_{i}^{\alpha }\tau _{i}^{\beta }$ $(\alpha ,\beta
=x,y,z)$ compose the fifteen generators for a SU(4) group, 
\begin{equation}
\{T_{i}^{m}\}=\{2{\bf S}_{i}^{\alpha },2\tau _{i}^{\beta },4{\bf S}%
_{i}^{\alpha }\tau _{i}^{\beta }\},(m=1,2,\cdots ,15)
\end{equation}
with $\sum_{\gamma }T_{i}^{m}T_{i}^{m}=15$ and $P_{ij}=\sum_{\gamma
}(T_{i}^{m}T_{j}^{m}+1)/4.$ To explore the physical meaning of $P_{ij},$ we
define four possible states $\left| i\mu \right\rangle $ on each lattice
site according to the eigenvalues of $s^{z}$ and $\tau ^{z}$, \ where $\mu
=\left( s^{z},\tau ^{z}\right) $ or simply1, 2, 3 and 4.\ \ Define the total
SU(4) spin $T_{tot}=\sum_{i}T_{i}.$ Due to the symmetry of the model we have 
$[H,T_{tot}]=0.$\ The total SU(4) spin $T_{tot}$ is a good quantum number.
The operator $P_{ij}$ is in fact a permutation operator when it is applied
on the state $\left| i\mu ,j\nu \right\rangle $ 
\begin{equation}
P_{ij}\left| i\mu ,j\nu \right\rangle =\left| i\nu ,j\mu \right\rangle
\label{permutation}
\end{equation}
with $P_{ij}^{2}=1,$ where we have used the standard relation for spin 1/2
system.\cite{Hulthen38} The two eigenvalues of $P_{ij\text{ }}$is $\pm 1.$
This gives an upper and lower bound for energy per bond in Eq. (\ref{model}%
), i.e., $-J_{ij}\leq J_{ij}\left\langle P_{ij}\right\rangle \leq J_{ij}$
for any state$.$ For a two-site problem, there are six eigenstates for $%
P_{ij}$ with eigenvalue -1, $\left( \left| i\mu ,j\nu \right\rangle -\left|
i\nu ,j\mu \right\rangle \right) /\sqrt{2}$ where $\mu \neq \nu .$ The total 
$\left( T_{i}+T_{j}\right) ^{2}=20$, which indicates that a SU(4) singlet
cannot be formed at two sites. The minimal number of lattice sites to form
SU(4) singlet is four as shown by Li et al.\cite{Li98} An SU(4) singlet is
written as 
\[
su_{4}(i,j,k,l)=\sum_{\mu ,\nu ,\gamma ,\delta }\Gamma _{\mu \nu \gamma
\delta }\left| i\mu ,j\nu ,k\gamma ,l\delta \right\rangle 
\]
where $\Gamma $ is an antisymmetric tensor. Alternatively, denote spin and
orbital SU(2) singlets for sites i and j by $s(ij)$ and $\tau (ij)$,
respectively. An SU(4) singlet can be expressed in terms of spin and orbital
SU(2) singlets\cite{Bossche00} 
\begin{eqnarray}
su_{4}(i,j,k,l) &=&\sqrt{\frac{2}{3}}\left[ s(ij)s(kl)\tau (il)\tau
(jk)\right.  \nonumber \\
&&\left. -s(il)s(jk)\tau (ij)\tau (kl)\right] .
\end{eqnarray}
Exchange of the order of i, j, k, and l gives the same state. For any two
sites $i\prime $ and $j\prime $ among $i,j,k$, and $l$, we have 
\begin{equation}
P_{i^{\prime }j^{\prime }}su_{4}(i,j,k,l)=-su_{4}(i,j,k,l).  \label{su4}
\end{equation}
The exchange of the positions of four sites in the singlet keeps the singlet
unchanged. {\it Since -1 is the smallest eigenvalues of }$P_{ij}${\it , for
a four-site problem with all }$J_{ij}\geq 0${\it , the lowest energy state
is }$su_{4}(1,2,3,4)${\it \ with eigenvalues }$-\sum_{ij}J_{ij}.$ It is
worth noting that the conclusion is independent of the values of the
coupling $J_{ij}.$ Furthermore, by using Eq. (\ref{su4}), it is not hard to
check, 
\begin{equation}
\left( T_{i}+T_{j}+T_{k}+T_{l}\right) ^{2}su_{4}\left( i,j,k,l\right) =0.
\end{equation}
This identity indicates that the total SU(4) $T_{tot}=\sum_{i}T_{i}$ is
zero. There exists another important identity for two SU(4) singlets in
eight sites when indices $i_{1}$ and $j_{1}$ in $P_{i_{1}j_{1}}$ belong to
different singlets 
\begin{eqnarray}
&&P_{i_{1}j_{1}}su_{4}(i_{1},i_{2},i_{3},i_{4})su_{4}(j_{1},j_{2},j_{3},j_{4})
\nonumber \\
&=&su_{4}(j_{1},i_{2},i_{3},i_{4})su_{4}(i_{1},j_{2},j_{3},j_{4})
\label{2su4}
\end{eqnarray}
To prove the identity we utilize the permutation properties of $P$ as shown
in Eq.(\ref{permutation}). The resulting state is obtained by exchanging two
positions of $i_{1}$ and $j_{1}$ in different singlets.

To proceed further we introduce a concept of generalized RVB state. An SU(4)
RVB state is composed of SU(4) singlets, instead of SU(2) singlet. In
principle an SU(4) RVB state consists of all possible configuration, which
contain either the nearest neighbor SU(4) singlets or the long-range SU(4)
singlets. Depending on the Hamiltonian and the underlying lattice a SU(4)
RVB state as a ground state may have a different form. For instance in the
example we shall present later the state is a short-range RVB state. The
completeness of the RVB states as a basis for a global singlet state can be
shown from the group theory. Take the direct product of $N_{\Lambda }(=4M)$
states $\left| i\mu \right\rangle $ as basis. Young Tableaux is used to
represent the irreducible representation. If the irreducible representation
is a singlet the Young tableaux must be of the form of a $4\times M$
rectangle. In each column it is antisymmetrized, and in each row it is
symmetrized. In this way the Young tableaux represents a generalized RVB
state, as in the case for SU(2) system.\cite{Ma88} Since the irreducible
representation form a complete set, a linear combination of a RVB state is
another one. The number of the generalized RVB states are $(4M)!/(M!)^{4}.$
It is overcomplete and non-orthogonal. The Lanczos method can reorganize the
states to form a complete and orthogonal set of basis. Thus, we have the
following statement,

{\it Given that the number of lattice sites }$N_{\Lambda }=4M$ ($M$ is an
integer){\it , the SU(4) isotropic state of the symmetric spin-orbital model
with } {\it can be expressed as a linear combination of configurations
consisting four-site SU(4) singlets, i.e., SU(4)\ RVB state.}

Several remarks are in order.

(1). Following the Lanscoz method we can to re-construct the Hamiltonian in
a tridiagonal form on a set of complete and orthogonal basis by utilizing
Eqs.(\ref{su4}) and (\ref{2su4}). Each of the basis can be expressed in a
linear combination of SU(4) RVB\ states.

(2). In one-dimensional chain with 4M site the short-range SU(4) RVB state
composed by the nearest neighbor four-site SU(4) singlet has the energy per
bond -0.75J, which is very closed to the exact energy via Bethe ansatz
-0.82511J.\cite{Sutherland75} Numerical calculations on finite size lattice
suggest that the ground state on a square lattice may be a SU(4)\ RVB state. 
\cite{Bossche00}

(3). On a SU(2) antiferromagnetic model on a hypercubic lattice, it was
shown that the ground state is a spin singlet.\cite{Marshall55} We postulate
that this result is valid for SU(N) systems if the lattice can be decomposed
into N sublattices. Numerical calculations for finite clusters supports this
idea.

We now make use of the identities to construct two types of solvable models
with SU(4) symmetry. In principle, if we can write the Hamiltonian in the
form of the sum of semi-positive operators, and find a state which has
lowest eigenvalues for all semi-positive operators, the state must be the
ground state of the Hamiltonian. The method was used for spin 1/2 system by
Majundar and Ghosh.\cite{Majumdar70} The first type of solvable model is
defined on a d-dimensional hypercubic lattice. Label the lattice site by $%
i\in \Lambda .$ Each site contains four SU(4) spins. The SU(4) operators is
denoted by $T_{i\gamma }$. ($\gamma =1,2,3,4$). Assume the number of lattice
sites $i$ is $N.$ The total number of lattice sites is $4N.$ The model
Hamiltonian is 
\begin{eqnarray}
H &=&zJ^{\prime }\sum_{i,\gamma \neq \gamma ^{\prime }}\left( 2{\bf S}%
_{i\gamma }\cdot {\bf S}_{i\gamma ^{\prime }}+\frac{1}{2}\right) \left(
2\tau _{i\gamma }\cdot \tau _{i\gamma ^{\prime }}+\frac{1}{2}\right) 
\nonumber \\
&&+J\sum_{\left\langle i,j\right\rangle ,\gamma ,\gamma ^{\prime }}\left( 2%
{\bf S}_{i\gamma }\cdot {\bf S}_{j\gamma ^{\prime }}+\frac{1}{2}\right)
\left( 2\tau _{i\gamma }\cdot \tau _{j\gamma ^{\prime }}+\frac{1}{2}\right)
\label{solvable1}
\end{eqnarray}
where\ the intra-site coupling is large than the inter-site coupling $%
J^{\prime }=J(\alpha ^{2}+\frac{1}{\alpha ^{2}})/2\geq J$, z is the
coordinate number and $\alpha $ is an arbitrary number. To find the lowest
energy state, the Hamiltonian is rewritten as 
\begin{eqnarray*}
H &=&\frac{J}{16}\sum_{ij}\left[ \sum_{\gamma =1}^{4}\left( \frac{1}{\alpha }%
T_{i\gamma }+\alpha T_{j\gamma }\right) \right] ^{2} \\
&&-\sum_{\left\langle ij\right\rangle }(6/\alpha ^{2}+6\alpha ^{2}-4)J
\end{eqnarray*}
The Hamiltonian is semi-positive definite except for a constant. If we can
find a state $\left| \Phi \right\rangle $such that 
\[
\sum_{ij}J\left[ \sum_{\gamma =1}^{4}\left( \frac{1}{\alpha }T_{i\gamma
}+\alpha T_{j\gamma }\right) \right] ^{2}\left| \Phi \right\rangle =0, 
\]
this state must be the ground state. Here we construct an SU(4) valence bond
(VB) state 
\begin{eqnarray*}
\left| SVB\right\rangle
&=&su_{4}(i_{1}1,i_{1}2,i_{1}3,i_{1}4)su_{4}(i_{2}1,i_{2}2,i_{2}3,i_{2}4) \\
&&\cdots su_{4}(i_{N_{\Lambda }}1,i_{N_{\Lambda }}2,i_{N_{\Lambda
}}3,i_{N_{\Lambda }}4).
\end{eqnarray*}
It means that at each site $i$, four $T_{i\gamma }$ form an SU(4) singlet.
We can regard the state as an SU(4) singlet solid or VB solid at the
lattice. Therefore we have 
\[
\left[ \sum_{\gamma =1}^{4}\left( \frac{1}{\alpha }T_{i\gamma }+\alpha
T_{j\gamma }\right) \right] ^{2}\left| SVB\right\rangle =0. 
\]
for any pair of i and j. Alternatively, 
\begin{eqnarray*}
&&\left( \sum_{\gamma \neq \gamma ^{\prime }}\left( \frac{1}{\alpha ^{2}}%
P_{i\gamma i\gamma ^{\prime }}+\alpha ^{2}P_{j\gamma j\gamma ^{\prime
}}\right) +\sum_{\gamma ,\gamma ^{\prime }}P_{i\gamma j\gamma ^{\prime
}}\right) \left| SVB\right\rangle \\
&=&-(6/\alpha ^{2}+6\alpha ^{2}-4)\left| SVB\right\rangle .
\end{eqnarray*}
Hence $\left| SVB\right\rangle $ is the ground state of the model (Eq.(\ref
{solvable1})). The ground state energy per bond is $(6/\alpha ^{2}+6\alpha
^{2}-4)J.$ In this state there does not exist long-range correlation. The
short range RVB state is a typical quantum frustrated spin-orbital liquid.
When we want to break an SU(4) singlet it will cost a finite energy. Thus
the elementary excitation on this state has an energy gap.

The second type of solvable model is defined on a lattice which is
decomposed into two sublattices. The sublattice A labeled by \{j\} has one
SU(4) $T$ on each site. A lattice site belonging to sublattice B (labeled by
\{i\}) is located on the middle of two sites j. Each site i contains four $%
T_{i\gamma }$ ($\gamma =1,2,3,4$)$.$ The model Hamiltonian is defined as 
\begin{eqnarray}
H &=&2J\sum_{i,\gamma \neq \gamma ^{\prime }}\left( 2{\bf S}_{i\gamma }\cdot 
{\bf S}_{i\gamma ^{\prime }}+\frac{1}{2}\right) \left( 2\tau _{i\gamma
}\cdot \tau _{i\gamma ^{\prime }}+\frac{1}{2}\right)  \nonumber \\
&&+J\delta \sum_{\left\langle i,j\right\rangle ,\gamma }\left( 2{\bf S}%
_{i\gamma }\cdot {\bf S}_{j}+\frac{1}{2}\right) \left( 2\tau _{i\gamma
}\cdot \tau _{j}+\frac{1}{2}\right)  \label{solvable2}
\end{eqnarray}
Similarly, the Hamiltonian may be rewritten as 
\[
H=\frac{J}{8}\sum_{\left\langle ij\right\rangle }\left[ \sum_{\gamma
=1}^{4}T_{i\gamma }+\delta T_{j}\right] ^{2}-\frac{J}{2}\sum_{\left\langle
ij\right\rangle }\epsilon _{0} 
\]
with $0<\delta <1$ and $\epsilon _{0}=12-2\delta +15\delta ^{2}/4.$ We can
construct a state $\left| \Phi \right\rangle $ such that all four $%
T_{i\gamma }$ at site i form SU(4) singlets and $T_{j}$ at site $j$ are in
any state. We have

\begin{equation}
\left[ \sum_{\gamma =1}^{4}T_{i\gamma }+\delta T_{j}\right] ^{2}\left| \Phi
\right\rangle =15\delta ^{2}\left| \Phi \right\rangle .  \label{state2}
\end{equation}
The eigenvalue $15\delta ^{2}$ is the lowest energy of the squared operator
in Eq.(\ref{state2}) when $\delta <1$. The equation holds for any pair of i
and j. Therefore the state is also the ground state of the second-type of
solvable models (\ref{solvable2}). Its ground energy per bond $E_{g}=\left(
6-\delta \right) J.$ The state is highly degenerated since each site j has
four-fold degeneracy. The total degeneracy is $4^{N_{j}}$ where $N_{j}$ is
the number of lattice sites j. Among the degenerated states some are SU(4)
singlets, which can be expressed as SU(4) RVB states, and some are not.
Therefore\ due to the fact that the degeneracy is in power law of lattice
number the ground state should be a spin and orbital liquid.

For the SU(4) symmetric spin-orbital model, mathematically, we can write the
Hamiltonian in terms of the generators of SU(4) groups in the fermion
representation. It provides us a routine to generalize the main results in
this paper to the systems with SU(N) symmetry.\cite{Affleck85} In this sense
our main result can be regarded as a generalization of Anderson's RVB idea
to SU(N) system. The solvable models are simply modified in this way: the
site $i\gamma $ contains N SU(N) spins. On the ground state the SU(N) spins
at the N site form a SU(N) singlet. Hence we construct the two types of
solvable SU(N) models. The coupling coefficients should be modified slightly
according to the structures of different lattices.

In summary, we propose a generalized SU(4) RVB picture for spin-orbital
model. A state with global SU(4) singlet can be expressed as a SU(4)\ RVB
state. The idea is also generalized to systems with SU(N) symmetry. We
construct two types of solvable models, and evaluate the ground state
energies. One ground state is a SU(4) RVB solid, and another one is a
spin-orbital liquid.

The authors thank Fu-Chun Zhang and Michael Ma for their helpful
discussions. This work was supported by a RGC grant Hong Kong and a CRCG
grant of The University of Hong Kong.

\end{document}